\documentstyle[preprint,prabib,aps]{revtex}
\begin{document}
\draft
\title{Flux Tube Solutions in Kaluza-Klein Theory}
\author{V. Dzhunushaliev 
\thanks{E-mail address: dzhun@freenet.bishkek.su}}
\address{Dept. of Phys. VCU, Richmond, VA 23284-2000, USA \\
and \\ 
Theor. Phys. Dept. KSNU, 720024, Bishkek, Kyrgyzstan}
\author{D. Singleton
\thanks{E-mail address:das3y@maxwell.phys.csufresno.edu}}
\address{Dept. of Phys. CSU Fresno, 2345 East San Ramon Ave., 
Fresno, CA 93740-8031 USA}

\date{\today}

\maketitle

\begin{abstract} 
Spherically symmetric, vacuum solutions in 5D and 7D 
Kaluza-Klein theory are obtained. These solutions 
are flux tubes with constant cross-sectional size, located 
between (+) and (-) Kaluza-Klein ``electrical'' 
and ``magnetic'' charges disposed respectively 
at $r = \pm\infty$ and filled with constant Kaluza-Klein 
``electrical'' and ``magnetic'' fields. These objects 
are surprisingly similar to the flux tubes which form 
between two monopoles in Type-II superconductors and 
also the hypothesized color field flux tube that is
thought to form between two quarks in the QCD vacuum.
\end{abstract}
\newpage

\section{Introduction} 

From Einstein and Wheeler down to the present 
day wormholes (WH) have 
been one of the most intriguing objects
of study in general relativity. 
For example, if WHs exist in nature then what is 
the linear size of the 
mouth of the wormhole ? If the mouth is sufficiently 
large then one has 
a classical gravitational object which can carry matter and 
energy between different spacetime points \cite{mor} ! 
If the mouth size is small (approximately the Planck length
$\approx L_{Pl}$) then one has a gravitational object  
which could play an important role in spacetime foam, 
and may even serve as a geometrical model
for material particles
such as the electron \cite{ein}, \cite{wh}. 
Wheeler coined the name  ``charge without charge''
and ``mass without mass'' for such objects. 

In 4D Euclidean gravity WH solutions were obtained in \cite{gid}.
A full review of this problem can be found in \cite{vis2}.
In Ref. \cite{dzh1} a WH-like solution in multidimensional
gravity without any kind of matter was given. Our point of view
is that such vacuum WHs should be important
for Wheeler's ``charge without charge'' and
``mass without mass'' objects \cite{dzh3}. The flux tube 
solutions discussed in this paper represent a limiting 
case of the vacuum WH-like solutions, where the wormhole 
mouths are separated by an infinite distance, and the 
cross section size of the WH-like solution is 
constant. In addition, the string-like character of the 
flux tube solutions, as well as their similarity to 
the flux tube structures in Type II superconductors 
and the QCD vacuum in the presence of quarks, make 
them interesting objects of study.
 
\section{5D flux tube solution}
 
In this article we first examine the standard vacuum 5D
Kaluza-Klein theory. We take the form of the metric as
\begin{equation} 
ds^{2} = e^{2\nu (r)}dt^{2} - r_0^2e^{2\psi (r) - 2\nu (r)} 
(d\chi +  \omega (r)dt - n\cos \theta d\varphi )^2 -
dr^{2} - a(r)(d\theta ^{2} +
\sin ^{2}\theta  d\varphi ^2), 
\label{1} 
\end{equation} 
where $\chi $ is the 5$^{th}$ extra coordinate;
$r,\theta ,\varphi$ are $3D$  ``spherical-polar'' coordinates 
(the quotations marks indicate that these coordinates 
are not the same as the standard spherical-polar coordinates 
of flat space); $n$ is an integer; $r \in \{ -R_0 , +R_0 \}$.
The form of the metric and the limit of r
indicates that our metric is that for a WH
($R_0$ may be equal to $\infty$). As in Ref. \cite{gross} 
the presence of the $n \cos \theta d\varphi$ term indicates 
the possible presence of a ``magnetic'' charge in the 
solution. We consider 5D Kaluza-Klein theory as gravity
on the principal bundle with a U(1) structural group and
with ordinary 4D Einstein spacetime as the base space \cite{dzh2}.
In this kind of theory no quantities depend
on the 5$^{th}$ coordinate because the 5$^{th}$ dimension is a
symmetric space (the gauge group U(1)).

Using a $REDUCE$ package for symbolic calculation the 5D
Einstein equations for the metric given in Eq. (\ref{1}) are
\begin{eqnarray}
\nu '' + \nu'\psi' + \frac{a'\nu'}{a} -
\frac{1}{2} r_0^2 \omega '^2e^{2\psi - 4\nu} = 0, 
\label{2-1}\\ 
\omega '' - 4\nu'\omega' + 3\omega '\psi ' +
\frac{a'\omega '}{a} = 0,
\label{2-2}\\
\frac{a''}{a} + \frac{a'\psi '}{a} - \frac{2}{a} +
\frac{Q^2}{a^2}e^{2\psi - 2\nu} = 0,
\label{2-3}\\
\psi '' + {\psi '}^2 + \frac{a'\psi '}{a} -
\frac{Q^2}{2a^2}e^{2\psi - 2\nu} = 0,
\label{2-4}\\
\nu '^2 - \nu '\psi ' - \frac{a'\psi '}{a} +
\frac{1}{a} - \frac{a'^2}{4a^2} -
\frac{1}{4}r_0^2\omega '^2 e^{2\psi - 4\nu} -
\frac{Q^2}{4a^2} e^{2\psi - 2\nu} = 0
\label{2-5}
\end{eqnarray}
here $Q = nr_0$. We will find that this represents the 
Kaluza-Klein ``magnetic'' charge. 
A particularly simple solution of these equations is
\begin{eqnarray}
q = Q,
\label{3-1}\\
a = \frac{q^2}{2} = const,
\label{7-1}\\ 
e^{\psi} = e^{\nu} = \cosh \left( \frac{r\sqrt{2}}{q} \right), 
\label{7-2}\\ 
\omega = \frac{\sqrt{2}}{r_0}\sinh \left( \frac{r\sqrt{2}}{q} \right)
\label{7-3}
\end{eqnarray}
In order to define the Kaluza-Klein ``electrical'' field
of this solution we multiple Eq. (\ref{2-2}) by $4 \pi$
and rewrite it in the following way:
\begin{equation}
\left( \omega ' e^{3\psi - 4\nu} 4 \pi a \right)' = 0. 
\label{8}
\end{equation}
This can be compared with the normal 4D Gauss's Law
\begin{equation}
\left ( E_{4D} S\right )' = 0, 
\label{9} 
\end{equation}
where $E_{4D}$ is the 4D electric field and $S = 4 \pi r^2$ is the
area of the 2-sphere $S^2$. For the solution given in Eqs. (\ref{3-1} - 
\ref{7-3}), where $r^2$ is replaced by $a = const$, we find that the 
surface area of the flux tube solution is $S_{flux} = 4 \pi a$. Using 
this fact and comparing Eqs. (\ref{8}) and
(\ref{9}) we can identify $\omega ' e^{3\psi - 4\nu}$
as the Kaluza-Klein ``electric'' field:
\begin{equation}
E_{KK} = \omega ' e^{3\psi - 4\nu} = \frac{2}{r_0q} =
\frac{q}{r_0 a} = const. 
\label{10}
\end{equation} 
The Kaluza-Klein ``magnetic'' field can be derived as in 
Ref. \cite{gross}. The gauge field associated with the 
metric in Eq. (\ref{1}) has a $\varphi$ component as 
$A_{\varphi} = -r_0 n \cos \theta$. The Kaluza-Klein ``magnetic'' 
field is then found from ${\bf H}_{KK} = {\bf \nabla} \times {\bf A}$, 
where the curl is taken using the metric of Eq. (\ref{1}) 
and the solution of Eqs. (\ref{3-1} - \ref{7-3}). The resultant 
Kaluza-Klein ``magnetic'' field derived from this is 
\begin{equation} 
{\bf H}_{KK} = {1 \over a \sin \theta} 
\left( {\partial \over \partial \theta} (- r_0 n \cos \theta) \right) 
 {\hat {\bf r}} = \frac{r_0 n}{a}{\hat {\bf r}} 
 = \frac{Q}{a} {\hat {\bf r}}. 
\label{11} 
\end{equation} 
Thus, this solution is \underline{a flux tube}
with constant Kaluza-Klein ``electrical'' and
``magnetic'' fields. The tube-like geometry of 
this solution comes from the fact that $a(r) = const$ 
rather than $a(r) = r^2$. The metric function 
$a(r)$ can be thought of as the cross-sectional 
size of the space at a particular $r$. For the present 
solution the cross-sectional size of the space 
remains constant as $r$ is varied and one has a tube-like 
space. The unit vector ${\hat {\bf r}}$ then points along 
the axis of the tube. Similar flux tube 
like solutions were investigated by Davidson and 
Gedalin \cite{david}. The direction of both the 
``electric'' and ``magnetic'' fields is along the 
${\hat {\bf r}}$ direction ({\it i.e.} along the axis 
of the flux tube). The sources
of these Kaluza-Klein fields (5D ``electric''
and ``magnetic'' charges) are located at $r = \pm \infty$.
This permits us to view this solution as a 5D
``electric'' and ``magnetic'' dipole. 

It is interesting to compare the 5D solution of Eqs.
(\ref{3-1}) - (\ref{7-3}) with the following 4D, nonasymptotically
flat, electrogravity solution which was originally investigated
by Levi-Civita \cite{levi} and rediscovered in Ref. \cite{ber}:
\begin{eqnarray}
ds^2 &=& a^2\left (\cosh^2 \zeta dt^2 - d\zeta^2 - d\theta ^2 -
\sin ^2 \theta d\varphi ^2\right ),
\label{12-1} \\ 
F_{01} &=& \rho ^{1/2} \cos\alpha, \;\;\; \; \; \; \;
F_{23} = \rho ^{1/2}\sin\alpha,
\label{12-1a}
\end{eqnarray}
where
\begin{equation}
G^{1/2} a \rho ^{1/2} = 1.
\label{12-2} 
\end{equation}
$\alpha$ is an arbitrary constant angle; $a$ and
$\rho$ are constants defined by Eq. (\ref{12-2});
$G$ is Newton's constant ($c=1$, the speed of light);
$F_{\mu\nu}$ is the electromagnetic field tensor.
For $\cos\alpha = 1$ ($\sin\alpha = 1$)
one has a purely electric (purely magnetic) field. 
Since our 5D solution had $G_{55} = const$ the solutions of Eqs.
(\ref{3-1}) - (\ref{7-3}) and Eqs. (\ref{12-1})-(\ref{12-2})
are similar. However, our 5D solution had the condition
that $\alpha = \pi /4$ ($E_{KK} = H_{KK}$). This constraint
indicates a difference between the 5D and 4D cases.
Generally speaking 5D gravity is equivalent to
4D gravity + electromagnetic + scalar fields \cite{per}.
For our 5D solution we find that
as a consequence of the 
condition $E_{KK} = H_{KK}$ the 5D gravitational field
solutions are identical to the 4D gravitational
plus electromagnetic field solutions.

In Ref.\cite{gun} the 4D solution of Eqs.
(\ref{12-1})-(\ref{12-2}) was used to construct
a composite WH. At the center of this composite
WH the solution of Eqs. 
(\ref{12-1})-(\ref{12-2}) was matched to two Schwarzschild
solutions. A similar construction was carried out
in Ref.\cite{dzh3} with one distinction:
the center of the composite WH in \cite{dzh3} was a vacuum
solution of 5D gravity with $G_{5t} \neq 0$ (this
led to the appearance of the 5D ``electric'' field).
 
\section{7D flux tube solution} 
 
Next we examine 7D Kaluza-Klein theory where the gauge group 
of the fibre is now SU(2). We take our metric to be of the form 
\begin{equation} 
\label{12} 
ds^2 = e^{2 \nu (r)} dt^2  -r_0 ^2 \sum _{a=1} ^3 (\sigma ^a 
-A_{\mu} ^a (r) dx ^{\mu} ) ^2 - dr^2
- a(r) (d \theta ^2 + \sin ^2 \theta d \varphi ^2 ) 
\end{equation} 
This is a simplified form of the metric used in \cite{dzh1}. 
The one-forms $\sigma ^a$ are given by 
\begin{eqnarray} 
\sigma ^1 &=& {1 \over 2} ( \sin \alpha d \beta -\sin \beta 
\cos  \alpha d \gamma )
\label{13} \\
\sigma ^2 &=& -{1 \over 2} ( \cos \alpha d \beta + \sin \beta 
\sin \alpha d \gamma )
\label{14} \\ 
\sigma ^3 &=& {1 \over 2} ( d \alpha  + \cos \beta 
d \gamma ) 
\label{15} 
\end{eqnarray} 
where $0 \le \beta \le \pi$, $0 \le \gamma \le 2 \pi$, 
$0 \le \alpha \le 4 \pi$ are Euler angles for the SU(2) group. 
The potentials, $A_{\mu} ^a$, are chosen to have the 
following monopole like form 
\begin{eqnarray} 
\label{16} 
A_{\theta} ^a &=& {1 \over 2} (\sin \varphi ; -\cos \varphi ; 0 ) \\ 
\label{17} 
A_{\varphi} ^a &=& {\sin \theta \over 2} 
(\cos \varphi \cos \theta ; \sin \varphi \cos \theta ; -\sin \theta ) \\ 
\label{18} 
A_t ^a &=& v(r) (\sin \theta \cos \varphi ; \sin \theta \sin \varphi ; 
\cos \theta ) 
\end{eqnarray} 
Substituting the metric ansatz of Eq. (\ref{12}) 
into the 7D Einstein vacuum equations we find, again 
using a REDUCE package for symbolic manipulations, the 
following set of coupled equations 
\begin{eqnarray} 
\label{19} 
\nu {''} + {\nu '} ^2 + {a' \nu ' \over a} - {r_0 ^2 \over 2} 
e^{-2 \nu} {v'} ^2 &=& 0 \\ 
\label{20} 
\nu {''} + {\nu '} ^2 + {a{''} \over a} - {{a'} ^2 \over 2 a^2} 
- {r_0 ^2 \over 2} e^{-2 \nu} {v'} ^2 &=& 0 \\ 
\label{21} 
{a{''} \over a} + {a' \nu ' \over a} - { 2 \over a } 
+ {r_0 ^2 \over 4 a^2} &=& 0 \\ 
\label{22} 
{r_0 ^2 \over 6} e^{-2 \nu} {v '} ^2 - {2 \over r_0 ^2} 
-{r_0 ^2 \over 24 a^2} &=& 0 \\ 
\label{23} 
v {''} - v' \left( \nu '- {a' \over a} \right) &=& 0 
\end{eqnarray} 
Eq. (\ref {23}) can be integrated once to give 
\begin{equation} 
v' = {q \over r_0 a} e^{\nu} 
\label{24} 
\end{equation} 
where $q$ is an integration constant. Inserting this result 
back into Eqs. (\ref{19} - \ref{23}) one finds a solution 
which is almost identical to the 5D case given by Eqs. (\ref{7-1} - 
\ref{7-3}) 
\begin{eqnarray} 
a &=& {2 q^2 \over 7} = {r_0 ^2 \over 8} = const 
\label{25} \\
e^{\nu} &=& \cosh \left( {7 r \over 2 \sqrt{2} q} \right)  
\label{26} \\
v &=& {\sqrt{2} \over r_0} \sinh \left( {7 r \over 2 \sqrt{2} q} \right) 
\label{27} 
\end{eqnarray} 
By comparing Eq. (\ref{27}) and Eq. (\ref{24}) we find 
that the integration constant $q = \sqrt{7a /2}$. This 
constant $q$ is the electric ``charge'' of the present 
solution. This 7D solution also has color ``electric'' 
and ``magnetic'' fields similar to those of the 5D 
solution. To see this most directly one can apply 
the following gauge transformation to the potentials of 
Eqs. (\ref{16} - \ref{18}) 
\begin{equation} 
\label{28} 
A'_{\mu} = S^{-1} A_{\mu} S - i (\partial _{\mu} S^{-1} )S 
\end{equation} 
where 
\begin{equation} 
\label{29} 
S =  \left ( 
\begin{array} {cc} \cos {\theta \over 2} & 
-e^{-i \phi} \sin {\theta \over 2} \\ 
e^{i \phi} \sin {\theta \over 2} & 
\cos {\theta \over 2}
\end{array} \right ) 
\end{equation} 
with this gauge transformation we find that the gauge 
potentials become 
\begin{eqnarray} 
\label{30a} 
A'^a _{\theta} &=& (0; 0 ; 0) \\ 
\label{30} 
A'^a _{\varphi} &=& (\cos \theta -1 ) (0 ; 0; 1) \\ 
\label{31} 
A'^a _t &=& v(r) (0; 0 ; 1) 
\end{eqnarray} 
So that only the $\sigma ^3$ direction in isospin space 
is non-zero. The calculation of the Kaluza-Klein ``magnetic'' 
and ``electric'' fields is now the same as for the 5D case 
since the ``Abelian'' gauge transformation 
given by Eqs. (\ref{28}) - (\ref{29})
brings us to a gauge where the non-Abelian fields take on 
an Abelian form. Putting back the factor of $r_0$ from the metric 
of Eq. (\ref{12}) in 
the gauge potentials we find that the Kaluza-Klein 
``magnetic'' field for the 7D case becomes 
\begin{equation} 
\label{32} 
{\bf H}_{KK} = H_{KK} ^3 = {1 \over a \sin \theta} 
\left( {\partial \over \partial \theta} (r_0 A^3 _{\varphi}) \right) 
{\hat {\bf r}} = \frac{- r_0}{a}{\hat {\bf r}} 
\end{equation} 
(Compared to the 5D case, we have from 
the outset specialized to $n=1$ for the 7D case). 
The Kaluza-Klein ``electric'' field is given by 
\begin{equation} 
\label{33} 
{\bf E}_{KK} = E_{KK} ^3 = {1 \over e^{\nu}} 
\left( {\partial \over \partial r} (r_0 A^3 _t) \right) {\bf {\hat r}} 
= {7 \over 2 r_0 q} {\bf {\hat r}} = {q \over r_0 a} {\bf {\hat r}} 
\end{equation} 
Thus, just as in 5D Kaluza-Klein theory, 7D Kaluza-Klein theory 
yields an infinite length flux tube solution. The constant 
cross section size of this solution is set by the constant, $a$. 
Also, as in the 5D case, the length of this tube is filled with 
uniform ``electric'' and ``magnetic'' fields. The 7D ``electric'' 
and ``magnetic'' charge which produce these fields are taken to be 
located at $r = \pm \infty$. 
 
\section{Discussion} 
 
We would like to emphasize the important role that the 
``magnetic'' charge plays in the formation of these flux tubes.
If the ``magnetic'' charge is zero ({\it i.e.} if $n=0$) then we
would have a wormhole-like solution located between two
null surfaces as in \cite{dzh1} \cite{dzh2}. The addition 
of ``magnetic'' charges  results in the wormhole-like 
solutions becoming the flux tube solutions presented here. 
This situation has similarities to the formation of flux tubes in Type II
superconductors. When a magnetic monopole and anti-monopole
are placed within a Type II superconductor a flux tube will
form between them. 
Also, in the dual superconductor picture of confinement 
for QCD, color ``electric'' monopoles are necessary in order that
a color field flux tube form between two color ``electric'' 
charges. One difference is that in these cases the flux tube is 
usually only filled with ``magnetic'' or ``electric'' fields. 
For the solutions presented here both ``electric'' and 
``magnetic'' fields occur within the flux tube. 

The asymptotic behaviour of this flux tube
solution is interesting. At infinity the time-time 
part of the 5D metric approaches
$G_{tt} = \exp{(2\nu)}\stackrel{r\rightarrow\pm\infty} 
{\longrightarrow}\infty$. This situation is easy to
understand: at infinity the area of the $S^2$ sphere is
$4\pi a$ rather than $4\pi r^2$. This means that our
solution is not an asymptotically flat wormhole. This indicates 
that this solution, while probably having limited use as 
a macroscopic object as in \cite{mor}, may be of interest
as a Planck scale object in connection with ideas on 
spacetime foam or as a geometrical model for material 
objects such as electrons or strings.
 
\section{Acknowledgements} This work has been funded by the 
National Research Council under the Collaboration
in Basic Science and Engineering Program.

\end{document}